# Charge of a quasiparticle in a superconductor

*Yuval Ronen*[†,1], *Yonatan Cohen*[†,1], *Jung-Hyun Kang*[1], *Arbel Haim*[1], *Maria-Theresa Rieder*[1,2], *Moty Heiblum*[#,1], *Diana Mahalu*[1] *and Hadas Shtrikman*[1]

[1]*Braun Center for Submicron Research, Department of Condensed Matter Physics, Weizmann Institute of Science, Rehovot* 76100, *Israel*

[2]*Dahlem Center for Complex Quantum Systems, Freie University, Berlin* 14195, *Germany*

[†] *Equal contributions*

[#]*Corresponding Author* (*moty.heiblum@weizmann.ac.il*)

**Abstract**

**Non-linear charge transport in SIS Josephson junctions has a unique signature in the shuttled charge quantum between the two superconductors. In the zero-bias limit Cooper pairs, each with twice the electron charge, carry the Josephson current. An applied bias $V_{SD}$ leads to multiple Andreev reflections (MAR), which in the limit of weak tunneling probability should lead to integer multiples of the electron charge $ne$ traversing the junction, with $n$ integer larger than $2\Delta/eV_{SD}$ and $\Delta$ the superconducting order parameter. Exceptionally, just above the gap $eV_{SD} \geq 2\Delta$, with Andreev reflections suppressed, one would expect the current to be carried by partitioned quasiparticles; each with energy dependent charge, being a superposition of an electron and a hole. Employing shot noise measurements in an SIS junction induced in an InAs nanowire (with noise proportional to the partitioned charge), we first observed quantization of the partitioned charge $q = e^*/e = n$, with $n$=1- 4; thus reaffirming the validity of our charge interpretation. Concentrating next on the bias region $eV_{SD} \sim 2\Delta$, we found a reproducible and clear dip in the extracted charge to $q \sim 0.6$, which, after excluding other possibilities, we attribute to the partitioned quasiparticle charge. Such dip is supported by numerical simulations of our SIS structure.**

Excitations in superconductors (Bogoliubov quasiparticles) can be described according to the BCS theory (Bardeen–Cooper-Schrieffer) [1], as an energy dependent superposition of an electron with amplitude $u(\varepsilon)$, and a hole with amplitude $v(\varepsilon)$; where the energy $\varepsilon$ is measured relative to the Fermi energy [2]. Evidently, the expectation value of the charge operator (applied to the quasiparticle wave-function), which we address as the quasiparticle charge $e^*=q(\varepsilon)e$, is smaller than the charge of an electron, $q(\varepsilon) = |u(\varepsilon)|^2 - |v(\varepsilon)|^2$ [3]. Solving the Bogoliubov-de Gennes equations one finds that $|u(\varepsilon)|^2 = \frac{1}{2}\left[1+\frac{\sqrt{\varepsilon^2-\Delta^2}}{\varepsilon}\right]$ and $|v(\varepsilon)|^2 = \frac{1}{2}\left[1-\frac{\sqrt{\varepsilon^2-\Delta^2}}{\varepsilon}\right]$, with the expected charge evolving with energy according to $q(\varepsilon) = \frac{\sqrt{\varepsilon^2-\Delta^2}}{\varepsilon}$ - vanishing altogether at the superconductor gap edges [3]. Note, however, that the quasiparticle wave-function is not an eigen-function of the charge operator [3, 4]. Properties of quasiparticles, such as the excitation spectra [5], lifetime [6-10], trapping [11], and capturing by Andreev bound states [12, 13], had already been studied extensively; however, studies of their charge is lagging. In the following we present sensitive Shot noise measurements in a Josephson junction, resulting in a clear observation of the quasiparticle charge being smaller than $e$, $q(eV_{SD}\sim 2\Delta)<1$, and evolving with energy, as expected from the BCS theory.

In order to observe the BCS quasiparticles in transport we study a Superconductor-Insulator-Superconductor (SIS) Josephson junction in the non-linear regime. The overlap between the wave functions of the quasiparticles in the source and in the drain is expected to result in a tunneling current of their effective charge. This is in contrast with systems which are incoherent [14, 15] or with an isolated superconducting island, where charge conservation leads to traversal of multiples of $e$ – Coulomb charge [16]. As current transport in the non-linear regime results from 'multiple Andreev reflections' (MAR), it is prudent to make our measurements credible by first measuring the charge in this familiar regime.

In short, the MAR process, described schematically in Fig. 1, carries a signature of the shuttled charge between the two SCs, being a consequence of $n$ traversals through the junction (as electron-like and hole-like quasiparticles), with $n$ an integer larger than $2\Delta/eV_{SD}$. A low transmission probability $t$ (via tunneling through a barrier) in the bias range $2\Delta/n < eV_{SD} < 2\Delta/(n-1)$ assures dominance of the lowest order MAR process (higher orders are suppressed as $t^n$); with the charge evolving in nearly integer multiples of the electron charge. While there is already a substantial body of theoretical [3, 17-23] and experimental [24-29] studies of the MAR process, charge determination without adjustable parameters is still missing. An important work by Cron *et al*. [27] indeed showed a staircase-like behavior of the charge using 'metallic break-junctions'; however, limited sensitivity and the presence of numerous conductance channels, some of which with relatively high transmission probabilities, did not allow exact charge quantization. Our shot noise measurements, performed on a quasi-1D Josephson junction (single mode nanowire) allowed clear observation of charge quantization without adjustable parameters. To count a few advantages: (*i*) The transmission of the SIS junction could be accurately controlled using a back-gate; (*ii*) This, along with our high sensitivity in noise measurement, enabled us to pinch the junction strongly (thus suppressing higher MAR orders); and (*iii*) With the Fermi level located near the 1D channel van Hove singularity, a rather monoenergetic distribution could be injected (supplementary section: S7).

Our SIS Josephson junction was induced in a back-gate controlled, single channel nanowire (NW). The structure, shown in Fig. 2, was fabricated by depositing two Ti/Al (5nm/120nm) superconducting electrodes, 210nm apart, onto a bare ~50nm thick InAs NW, baring a pure wurtzite structure, grown by the gold assisted vapor-liquid-solid (VLS) MBE process. The Si:P$^+$ substrate, covered by SiO$_2$ (150nm thick), served as a back-gate (BG), allowing control of the number of conducting channels in the NW (S2). While the central part of the NW could be fully depleted, the segments intimately covered by the Ti/Al superconducting electrodes are flooded with carriers and are barely affected by the back-gate voltage. The density therefore decreases smoothly towards the depleted region in the very center of the junction, so that the actual tunnel barrier is much narrower than 200nm. On the other hand, the induced SC coherence length is expected to be much larger than 200nm - assuring coherence of electron-hole quasiparticles along the junction [30].

Since the probability of each single-path MAR process is $t$, and the probability for $n$ paths scales as $t^* = t^n$, a sufficiently small $t$ is necessary to single out the most probable (lowest $n$) MAR process. This evidently leads to a minute shot noise signal, requiring sensitive electronics and weak background noise. A 'cold' (~1K), low noise, homemade preamplifier was employed, with a sensitivity limit better than ~$10^{-30}$A$^2$/Hz at ~600kHz. Interested in the current dependent 'excess noise' (with spectral density $S^i_{exc}$), the non-shot noise components should be recognized and subtracted. The latter are composed of a thermal (Johnson-Nyquist) component, $4k_BTr$ [31,32], the preamplifier's current noise $S^i_{amp}$ (current fluctuations driven back to our device), and its voltage noise $S^v_{amp}$; while the ubiquitous 1/$f$ noise (due to multiple sources) is negligible at our measurement frequency (S4). Altogether $S^v(0)$ is given by:

$$S^v(0) = S^i_{exc}(0)r^2 + 4k_BTr + S^i_{amp}(0)r^2 + S^v_{amp} \quad , \qquad (1)$$

where $k_BT$ is the thermal energy and $r$ is the total resistance of the SNS junction and the load resistance, namely, $R_{sample}$+5Ω in parallel with $R_L$ (see Fig. 2). Hence, the 'zero frequency excess noise' for a stochastically partitioned single quantum channel at sufficiently low temperature (our $k_BT$~2μeV while $eV_{SD}$=50-300μeV) [33-36] is:

$$S_{exc}(0) = 2e^*I(1-t^*) \quad , \qquad (2)$$

with $e^*=qe$, $I$ the net DC current, and $t^* = \dfrac{G}{qg_Q}$, where $g_Q = \dfrac{2e^2}{h}$ is the quantum conductance of a spin degenerate channel in the normal part of the wire (S6). Hence, the charge (in units of the electron charge, $e$) is:

$$q = \dfrac{S_{exc}(0)}{2eI} + \dfrac{G}{2e^2/h} \quad . \qquad (3)$$

Two comments regarding Eq.3 are due here: (*i*) Using the differential conductance $G$ for calculating the transmission probability at energies near $eV_{SD}$ is justified since most of the current is carried by quasiparticles emitted in a narrow energy window; much narrower than Δ due to the van Hove singularity of the density of states in the 1D NW (see more in the discussion part); and (*ii*) When the transmission is small so that $G/g_q$~0, one resorts to the familiar Schottky (Poissonian) expression of a classical shot noise [37].

While details of the measurement setup and the algorithm used in determining the true excess noise and the extracted charge are provided in the S3 and S4; a short description is due here. As seen in Fig. 2, conductance and noise were measured in the same configuration at an electron temperature of ~25mK. The differential conductance was measured by applying 1μV at 600kHz in addition to a variable DC bias, while noise was measured with an applied DC bias only. A load resistor of either $R_L$=1kOhm or $R_L$=20kOhm, shunted by a resonant *LC* circuit (with a center frequency of 600kHz), terminated the circuit to ground. The signal was amplified by a cascade of 'cold' and 'warm' amplifiers, and measured by a spectrum analyzer with an appropriate bandwidth. The smaller $R_L$ was used when the sample's resistance was relatively small, thus restraining fluctuations in the background noise on the bare sample fluctuating conductance. It is important to note that the use of a 'voltage source' for $V_{SD}$ (rather than a 'current source') allowed access to quiescent regions of negative differential conductance, which otherwise would have been inaccessible (being within hysteretic loops in the $I$-$V_{SD}$ characteristics).

We start with $R_L$=1kOhm and junction conductance tuned by the back-gate to a partly transmitted single channel in the bare part of the NW. Four MAR conductance peaks were observed at $V_{SD}$=2Δ/$en$=300μV/$n$ (note that the induced gap in the InAs NW is nearly that of the Al superconductor). The static $I$-$V_{SD}$ characteristic, required for the determination of the energy dependent charge, was obtained by integrating the differential conductance (Fig. 3b). After a careful subtraction of the background noise, (S4), we extracted the charge as shown in Fig. 3c. Clear steps are seen at values of $q=n$, with $1 \leq n \leq 4$. Higher charge values (for $n > 4$) are averaged out mostly due to the successively narrowing MAR region as ~$1/n^2$ and possibly some inelastic scattering events. It should be stressed out here that while the conductance (and thus the deduced $t^*$) and the total noise fluctuate violently, the extracted charge evolves smoothly between each of the quantized charge values – reassuring the process of charge extraction.

We performed numerical simulations of the conductance and the excess noise at various junction transmission coefficients and energies, with a Fano factor defined as $F=S_{exc}(0)/2eI$ (S1). The results shown in Fig. 3d, contrary to the experimental results in Fig. 3c, predict integer charge plateaus at much lower transmission probabilities $t\sim0.05$. We attribute this difference to the sharp density of states profile resulting from the position of the Fermi level near the van Hove singularity of the 1D nanowire alluded above [38] (supplementary section: S7), which suppresses higher orders MAR - thus allowing charge quantization at relative high transmissions. The vicinity of the Fermi level to the bottom of the conduction band was not taken into account in the theoretical model (S1 and S8). Consequently, the normalized excess noise $S^*_{exc} = S_{exc}/(1-t^*)$, plotted as a function of the current in Fig. 3e, reveals straight lines with quantized slopes, all crossing the origin, confirming that in each relevant bias regime quasiparticles indeed emerge within a narrow energy window.

The robust quantized plateaus of the extracted charge (in two different NWs) paved the way to the determination of the traversing charge near the superconductor gap edge. Singling out the $n=1$ process (having $t^*=t$), very close to $eV_{SD}=2\Delta$ requires strong suppression of the $n=2$ process ($t^*=t^2$); thus further increasing the barrier in the bare part of the NW, as evident by the weaker MAR processes in Fig. 4a and higher junction resistances (now $R_L=20$kOhm). A few $I$-$V_{SD}$ characteristics, obtained by integrating the differential conductance for several back-gate voltages, are plotted in Fig. 4b. The extracted traversing charge as a function of bias is shown in Fig. 4c for a few values of the transmission coefficient; with a clear dip in the charge appearing near $eV_{SD}=2\Delta$ for lower transmissions. In Fig. 4d we plot the lowest charge measured at each transmission probability $t$ – observing a minimum of $q\sim0.6$ at $t\sim0.05$. As the barrier is increased even further ($t<0.01$), the extracted charge increased towards $e$.

The numerical calculations results for the Fano factor, $F=S_{exc}(0)/2eI$, around $eV_{SD}\sim 2\Delta$ are shown in Fig. 4e & 4f for various transmissions. The theoretical calculation also resulted in a dip which emerges as the transmission is lowered similarly to our experimental data. The discrepancy in the values of the transmission, in which the dip appears, can once again be attributed to the sharper profile of the density of states. Another difference from the theoretical calculation that should be noted is the decrease in the apparent charge from $2e$ at $eV_{SD}<2\Delta$. We relate this decrease to unavoidable processes of charge $e$ transport which are of order $t$ (not $t^2$); such as quasiparticles excited by noise or temperature, and sub-gap current originating due to the soft induced gap.

In order to further test the validity of the dip in the extracted charge, we fabricated and tested a Normal-I-S (NIS) junction. Here too, a conductance peak develops at the gap's edge (this time at $eV_{SD}=\Delta$); however, the charge evolves monotonically from $e$ to $2e$, without any sign of a dip (Fig. 5a). This result is backed by our numerical simulations (Fig. 5b), while in S9 we give a more intuitive physical picture that reflects why charge partition should not be observed in NIS junction.

Our assertion of observing the quasiparticle charge near the gap's edge requires a discussion. One may consider three possible models of single quasiparticles transport near $eV_{SD}=2\Delta$: (*i*) The electric field may rip off each quasiparticle to its electron and hole components, thus accelerating only one component (say, electrons) towards the drain; leading to current noise of partitioned charges of $e$ and a Fano factor of 1 at $t\ll 1$. This might play a role in an SNS junction, but less likely in our SIS tunneling junction. (*ii*) In an SIS junction, the electric field across the insulating barrier (I) realigns full quasiparticle states in the source (S) with empty quasiparticle states in the drain (S), making tunneling events possible. One possibility is that each tunneling event collapses in the drain to an electron with probability $u^2$ or to a hole with probability $v^2$. In this case the expected charge fluctuations for $t\ll 1$ lead to a Fano factor $F=(u^2-v^2)^{-1}>1$ [3, 4] (see S7); (*iii*) Alternatively, each tunneling event is that of a coherent superposition of an electron and a hole, leading to a Fano factor $F=(u^2-v^2)<1$ at $t\ll 1$ (S7). Thus measuring a charge which is smaller than $e$ confirms the third scenario.

If fractionally charged quasiparticles indeed tunnel through the SIS junction, why does the extracted charge climb back to $e$ when the tunneling probability is extremely small? Specifically, an opaque barrier is expected to allow only tunneling of electrons, as both sides of the barrier should be quantized in units of the electronic charge due to charge neutrality (recall the similar findings in the FQHE [39, 40]).

In summary, employing sensitive, low frequency, shot noise measurements [41, 42, 43], we observed an evolution of energy dependent tunneling of quasiparticle charge in a SIS Josephson junction induced in a 'one-dimensional' InAs nanowire. The charge evolved as $e^*=ne$ away from the superconducting gap's edge, with $n=1$ for $eV_{SD}>2\Delta$ and $n=2$-$4$ for $eV_{SD}<2\Delta$ - in agreement with our understanding of multiple Andreev reflections (MAR). Moreover, at the gap's edge, $eV_{SD}\sim2\Delta$, with MAR processes strongly suppressed, the charge as inferred from the Fano factor was found to dip below the electron charge $e^*<e$; agreeing with the expectation value of the Bugoliubov quasiparticles being smaller than $e$. While such suppression of the Fano factor was observed by numerical simulations (Refs. 18 & 19 and here), the relation to the quasiparticle charge was so far never discussed. We suggest that this correlation between the suppressed shot noise and the quasiparticle charge in SIS junctions should be further investigated theoretically beyond the simplified theoretical picture. Moreover, similar measurements should be applied to less ubiquitous superconductors, such as topological p-wave superconductors or high-Tc superconductors, to investigate the nature of their quasiparticle excitations.

**Acknowledgements**

We thank Y. Oreg, K. Michaeli, E. Altman, O. Entin, A. Aharony, R. Ilan, A. Akhmerov, E. Zeldov, N. Ofek and H. Inoue for helpful discussions. We are grateful to Ronit Popovitz-Biro for professional TEM study of the NW's crystal structure. M.H. acknowledges the partial support of the Israeli Science Foundation (ISF), the Minerva foundation, the U.S.-Israel Bi-National Science Foundation (BSF), the European Research Council under the European Community's Seventh Framework Program (FP7/2007-2013)/ERC Grant agreement No. 339070, and the German Israeli Project Cooperation (DIP). H.S. acknowledges partial support by ISF grant number 532/12, and IMOST grants #0321-4801 & #3-8668. H.A. is supported partially by ISF, grant number 532/12, the Minerva foundation, the European Research Council under the European Community's Seventh Framework Program (FP7/2007-2013)/ERC Grant agreement No. 340210. M.T.R. is supported by the Alexander von Humboldt Foundation in the framework of the Alexander von Humboldt Professorship, endowed by the German Federal Ministry of Education and Research.

**Figure Captions**

**Figure 1. Multiple Andreev Reflection (MAR)**. Illustrations of the leading processes contributing to the current as function of bias. In general, for $\frac{2\Delta}{n-1} > eV_{SD} > \frac{2\Delta}{n}$ the leading charge contribution to the current is *ne*. An electron-like quasiparticle is denoted by a full circle, while a hole-like quasiparticle is denoted

by an empty circle. (a) When the bias is larger than the energy gap, $eV_{SD} > 2\Delta$, the leading process is a single-path tunneling of single quasiparticles from the full states (left) to the empty states (right). This current is proportional to the transmission coefficient $t$. Higher order MAR process (dashed box), being responsible for tunneling of Cooper pairs, is suppressed as $t^2$. (b) For $2\Delta > eV_{SD} > \Delta$, the main charge contributing to the current is $2e$ with probability $t^2$. (c) For $\Delta > eV_{SD} > \frac{2\Delta}{3}$, the main charge contributing to the current is $3e$ with probability $t^3$.

**Figure 2. Scanning electron micrograph of the device and the circuit scheme.** InAs NW contacted by two superconducting Al electrodes. Conductance measurement: Sourcing by AC+DC, $V_{AC}$=0.1µV at ~600kHz, with AC output on $R_L$. Noise measurement: Sourcing by DC and measuring voltage fluctuations on $R_L$ at a bandwidth of 10kHz.

**Figure 3. Shuttled charges in the MAR process. (a)** Differential conductance (in units of $e^2/h$) as a function of applied bias, $V_{SD}$, normalized by $\Delta/e$, where $\Delta = 150\mu eV$ is the superconducting order parameter. The signature of the MAR processes is manifested by a series of peaks in bias corresponding to $\frac{eV_{SD}}{\Delta} = \frac{2}{n}$. **(b)** The $I$-$V$ characteristics as obtained by integrating the differential conductance. **Inset**: A zoom of the small current range. **(c)** The shuttled charge $q$ determined from Eq. (3) plotted as a function of $eV_{SD}/\Delta$. The pronounced staircase demonstrates the quantization of charge involved in the MAR processes. **(d)** Numerical simulations of the Fano factor, $F=S_{exc}/2eI$, as function of $eV_{SD}/\Delta$ for different values of the normal-region transmission $t$=0.4, 0.2, 0.1, 0.05 (the transmission at $eV_{SD} > 2\Delta$), according to S1. **(e)** The normalized excess noise (after dividing the excess noise by (1-$t^*$)), as a function of the current. Note that the local slope at every MAR region equals the global slope (red dashed curves; see also text and Eq. 3), suggesting a dominant contribution of a single process to the current and the noise near the energy corresponding to the bias. This in turn also suggests that most of the current originates

from a small energy range around the Fermi energy justifying the use of the differential conductance for extracting the transmission.

**Figure 4. Evolution of the quasiparticles charge near the edge of the gap. (a)** Differential conductance (in units of $e^2/h$) as a function of $eV_{SD}/\Delta$ for decreasing normal-region transmission $t=0.23, 0.15, 0.1$ (red, purple and blue respectively). As the transmission decreases (from blue to red) the conductance due to higher order processes diminishes with $t^n$ dependence. **(b)** The *I-V* curve obtained by integrating the differential conductance. **(c)** The charge determined from Eq. (3) plotted as a function of $eV_{SD}/\Delta$. As the transmission decreases, the value of the observed minima in the charge at the transition between $n=1$ and $n=2$ dips. **(d)** The measured charge $q$ is plotted as a function of the normal-region transmission $t$. **(e)** Results of numerical calculations showing $F=S_{exc}/2eI$ as function of $eV_{SD}/\Delta$ for low normal-region transmissions $t=0.2, 0.1, 0.05, 0.02, 0.01$. **(f)** Evolution of the minimum value of $F$ ($F_{min}$) as a function of transmission.

**Figure 5. Charge measurements in a Superconductor-Normal junction. (a)** The charge determined from Eq. (3) as a function of $eV_{SD}/\Delta$ at normal-region transmission $t=0.01$. The charge $q$ increases from 1 to 2 as $eV_{SD}$ crosses $\Delta$ **(b)** Numerical simulations of the Fano factor, $F=S_{exc}/2eI$, as a function of $eV_{SD}/\Delta$ at normal-region transmission $t=0.01$.

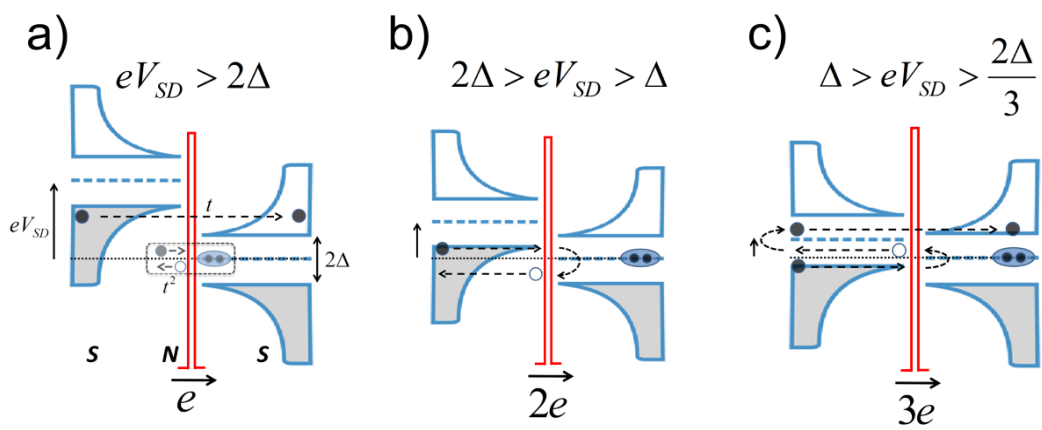

Fig. 1

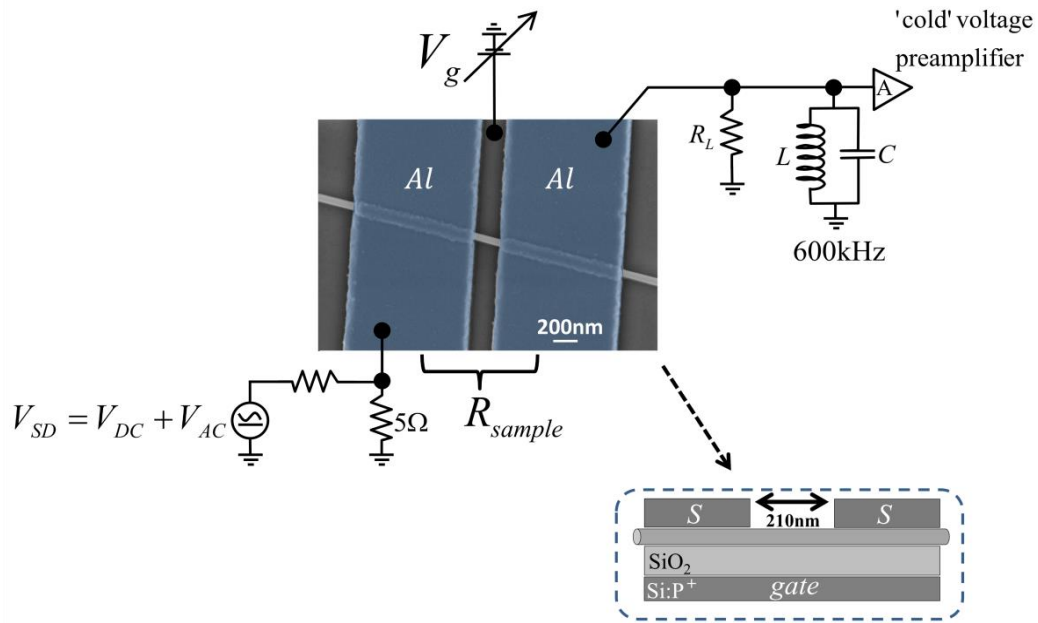

Fig. 2

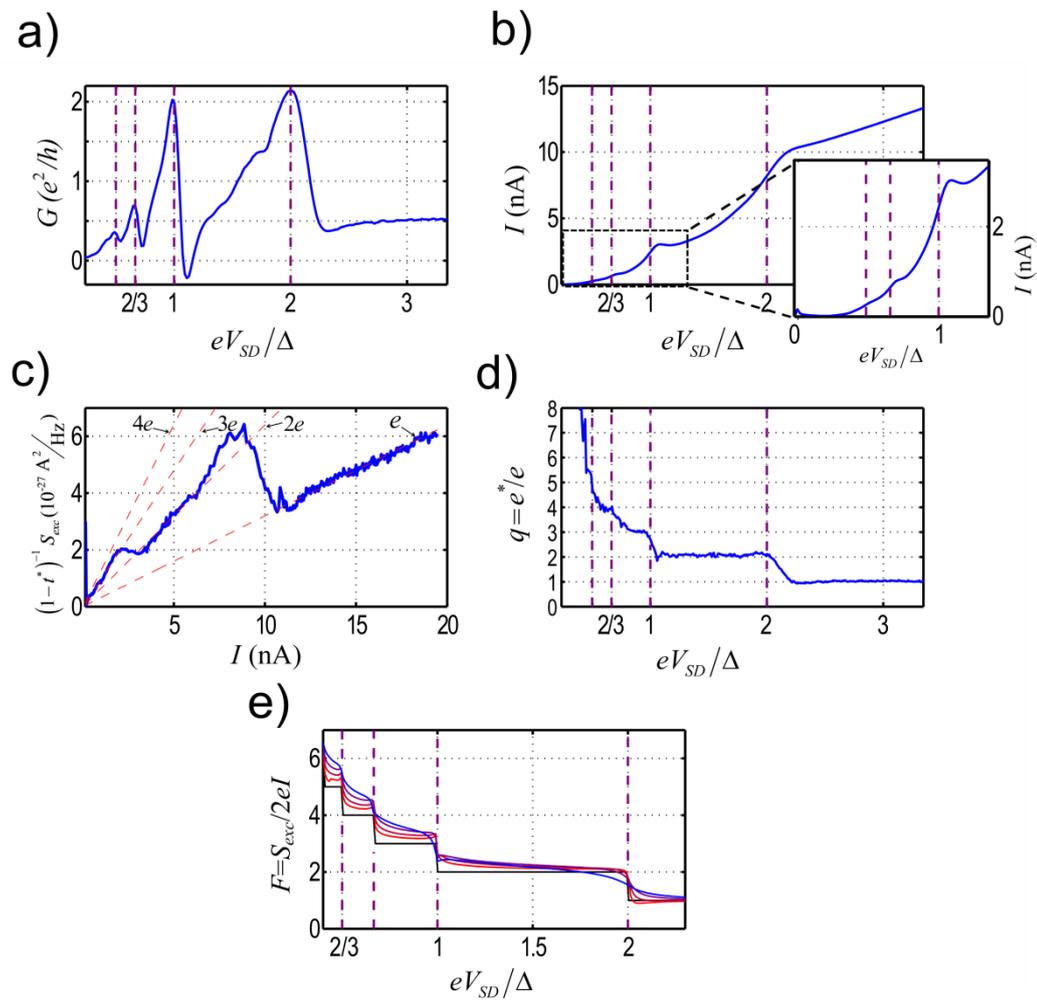

Fig. 3

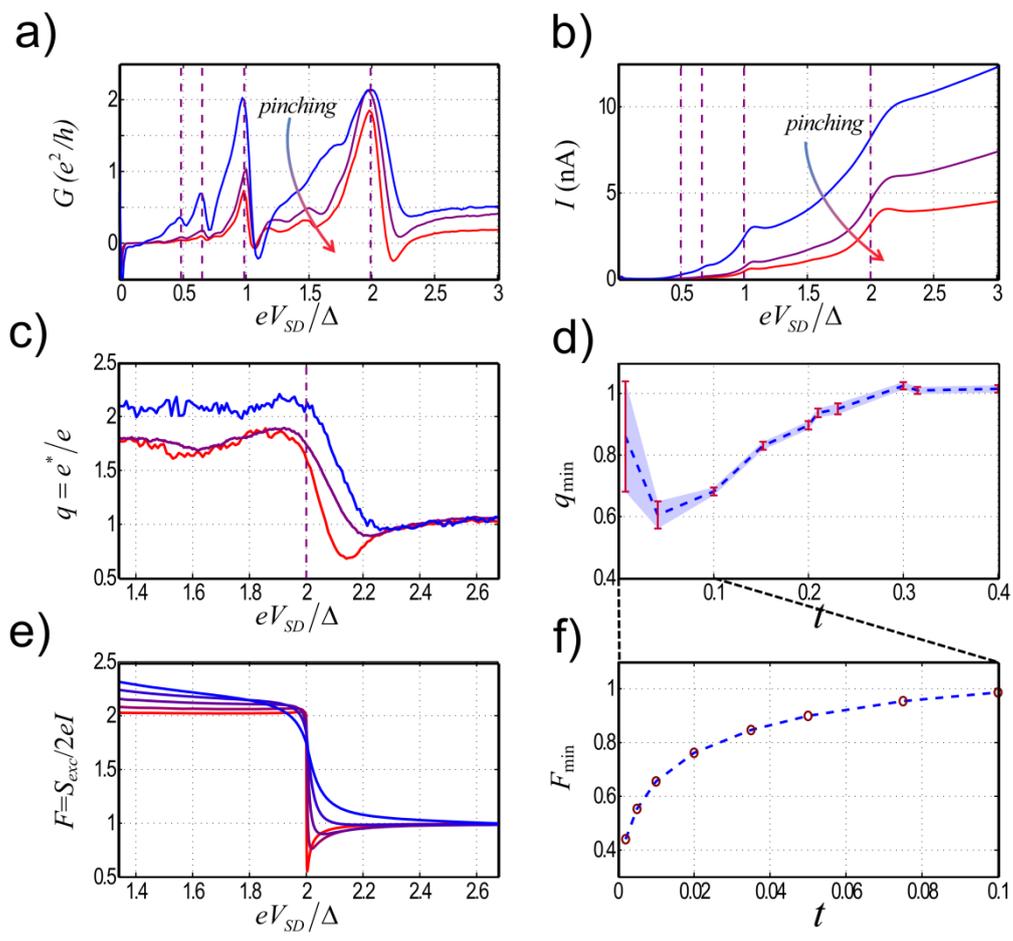

Fig. 4

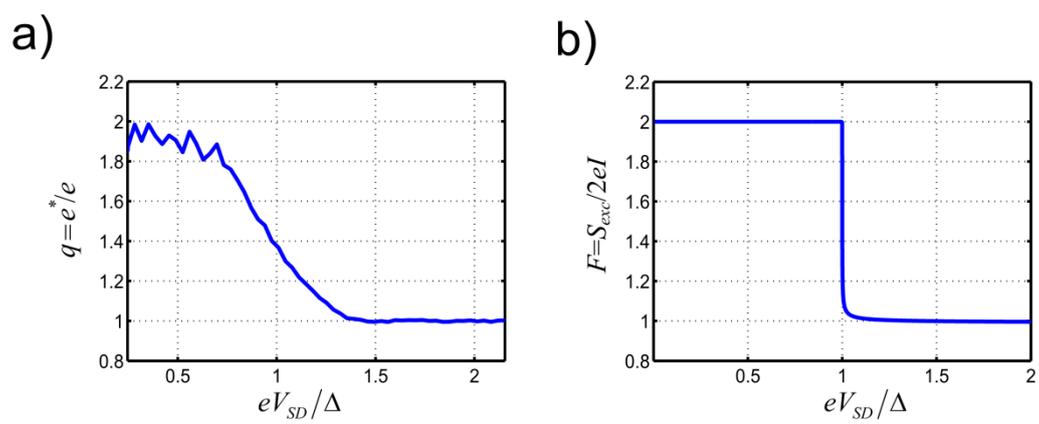

Fig. 5

# Charge of a quasiparticle in a superconductor

*Yuval Ronen[†,1], Yonatan Cohen[†,1], Jung-Hyun Kang[1], Arbel Haim[1], Maria-Theresa Rieder[1,2], Moty Heiblum[#,1], Diana Mahalu[1] and Hadas Shtrikman[1]*

[1]*Braun Center for Submicron Research, Department of Condensed Matter Physics, Weizmann Institute of Science, Rehovot* 76100*, Israel*

[2]*Dahlem Center for Complex Quantum Systems, Freie University, Berlin* 14195*, Germany*

[†] *Equal contributions*

[#]*Corresponding Author* (*moty.heiblum@weizmann.ac.il*)

## Methods and Supplementary Information

In this Supplementary Section we add details that could not find room in the main text. We placed a brief review of the theoretical background as well as the simulation. In addition, a few details of the NWs growth process followed by the fabrication process are provided, as well as more details on the conductance and noise measurements.

## S1 – Theoretical model

### Scattering theory of multiple Andreev reflections

Following [1, 2], we here outline the calculation of the current and current noise through a SNS-Josephson junction in the formalism of multiple Andreev reflections. Where electrons in the normal part are Andreev reflected from the superconducting leads. The normal region contains a barrier whose transmission amplitude squared is $t$. It is assumed that the length of the normal region is much smaller than the superconducting coherence length, and that the Fermi energy in the normal region is much larger than the superconducting gap $\Delta$. As a simplified setup we consider a short one-dimensional normal metal piece connected to one-dimensional semi-infinite superconducting leads.

A voltage biased Josephson junction exhibits a phase divergence that increases linearly in time as $\varphi(t) = 2eVt/\hbar$. This poses a periodically time-dependent problem which may be treated using Floquet theory expanding the eigenstates of the system in the quasi-energies $\varepsilon + 2neV$. A physical interpretation of these energies is in general not straightforward. However, in our specific case they are just the energies of the electrons/holes in the junction after multiple Andreev reflections. For instance, consider a voltage bias $eV \ll \Delta$ and an electron injected into the junction at energy $\varepsilon \leq -\Delta$ from the left lead- in equilibrium with a chemical potential $\mu$. After propagating through the junction this electron is Andreev reflected around the equilibrium potential of the right lead, $\mu - eV$, and returns as a hole with an energy $-\varepsilon - 2eV$. Here, it is again Andreev reflected into an electron with energy $\varepsilon + 2eV$. This process repeats until the particle has gained enough energy to overcome the superconducting gap and can be absorbed into the continuum of quasi-particle excitations in one of the leads.

Following the picture of multiple Andreev reflections, we can set up the wave-functions of the electronic state in the junction at the boundaries to the two leads, $x_1$, $x_2$ respectively. With the quasi-energies measured with respect to the chemical potential of the left lead, the wave-function for a quasi-particle incident on the junction from the left lead is

$$\phi(x_1,t) = \frac{J(\varepsilon)}{\sqrt{2\pi v_F}} \sum_{n=-\infty}^{\infty} \left[ \begin{pmatrix} a_{2n}A_n + \delta_{n,0} \\ A_n \end{pmatrix} e^{ikx_1} + \begin{pmatrix} B_n \\ a_{2n}B_n \end{pmatrix} e^{-ikx_1} \right] \cdot e^{-i(\varepsilon + 2neV)t}$$

$$\phi(x_2,t) = \frac{J(\varepsilon)}{\sqrt{2\pi v_F}} \sum_{n=-\infty}^{\infty} \left[ \begin{pmatrix} C_n \\ a_{2n+1}C_n \end{pmatrix} e^{ikx_2} + \begin{pmatrix} a_{2n+1}D_n \\ D_n \end{pmatrix} e^{-ikx_2} \right] \cdot e^{-i(\varepsilon + 2neV)t}$$

Where we chose a spinor in the basis $(c_\uparrow, c_\downarrow^\dagger)$ normalized to flux. The quasi-particle enters the junctions as an electron with probability $J(\varepsilon) = \sqrt{1 - |a(\varepsilon)|^2}$, with the Andreev reflection amplitude $a(\varepsilon) = (\varepsilon - i\sqrt{\Delta^2 - \varepsilon^2})/\Delta$ if $\varepsilon \leq \Delta$ and $a(\varepsilon) = (\varepsilon - sgn(\varepsilon)\sqrt{\Delta^2 - \varepsilon^2})/\Delta$ if $\varepsilon \geq \Delta$.

The tunneling barrier in the normal part of the junction with transmission $t \leq 1$ is implemented by a scattering matrix connecting the amplitudes of the left and the right side in a recursive fashion

$$\begin{pmatrix} B_n \\ C_n \end{pmatrix} = S \begin{pmatrix} a_{2n} A_n + \delta_{n,0} \\ a_{2n+1} D_n \end{pmatrix} \quad \text{and} \quad \begin{pmatrix} A_n \\ D_{n-1} \end{pmatrix} = S^* \begin{pmatrix} a_{2n} B_n \\ a_{2n-1} C_n \end{pmatrix}$$

$$S = \begin{pmatrix} r & d \\ d & -r^* d / d^* \end{pmatrix}$$

with $|d|^2 = t$.

The current $\hat{I}(t) = \dfrac{-ie\hbar}{2m} \sum_\sigma \left( \psi_\sigma^\dagger(t) \partial_x \psi_\sigma(t) - h.c. \right)$ may be evaluated at, say, the left boundary using the fact that each lead individually is in equilibrium. The electronic states are constructed from the Bogoliubov quasiparticle operators $\hat{\gamma}$ in the superconducting lead as $\psi_\sigma(t) = \sum_\nu \left( u_\nu(t) \hat{\gamma}_{\sigma,\nu} + sgn_\sigma u_\nu^*(t) \hat{\gamma}_{-\sigma,\nu} \right)$. We are using a joint index $\nu = i, \varepsilon$ to indicate the origin and energy of the incident electron ($i = l, r$ for the left and right lead). The wave-function $u_\nu(t) / \upsilon_\nu(t)$ is the respective electron/hole amplitudes obtained from solving the recursive relations denoted above. Our main interest in this work is the low-frequency current-noise, which is given by the zero Fourier component of the time averaged, symmetrized current correlation:

$$S(V) = \int d\tau \overline{\langle \delta \hat{I}(t) \delta \hat{I}(t+\tau) + \delta \hat{I}(t+\tau) \delta \hat{I}(t) \rangle}$$

where $\delta \hat{I}(t) = \hat{I}(t) - \langle \hat{I}(t) \rangle$, and the upper bar stands for time averaging (see [1, 2] for the explicit expressions). To comply with the experimentally measured quantity $S_{exc}(V)$, one subtracts from $S(V)$ the noise at zero voltage $S(V \to 0)$. For the experimentally-relevant temperatures and voltages this contribution is negligible.

**Details on the numerical calculations**

The solution to the recurrence relations described above is found using the method of continued fractions, following [3, 4]. First, the amplitudes $A_n, C_n, D_n$ are eliminated yielding a recursive equation for the $B_n$ of the general form

$$c_n B_{n+1} - d_n B_n + c_{n-1} B_{n-1} = -\delta_{n0}$$

Introducing a new variable $X_n = \frac{B_n}{B_{n-1}}$ for $n > 0$ and $X_n = \frac{B_n}{B_{n+1}}$ for $n < 0$, for a sufficiently large $n_{max} \gg 2\Delta/eV$. The physical reasoning of this ansatz is that an electron impinging a lead above the gap is absorbed with a probability approaching 1 rapidly for high energies. Hence, the amplitudes of states in the junction at high enough energies - corresponding to $n > n_{max}$ - are negligible. Following this procedure one can find all coefficients $B_n$ except for $B_0$, which is then obtained directly from Eq. (5) as

$$B_0 = \left(c_0 X_1 - d_0 + c_{-1} X_{-1}\right)^{-1}$$

**Additional simulations**

Noise in a NIS junction:

We here calculate the current and current noise in a junction of a normal metal (N) and a superconductor (S) with a tunneling barrier in the middle to model the insulating region (see Fig. S1). Transport through this kind of systems has been studied abundantly in the literature and we here adapt the formalism of Refs. [5, 6] to calculate the noise using the scattering matrix, particularly the reflection matrix with electron-hole grading

$$r(\varepsilon) = \begin{pmatrix} r_{ee}(\varepsilon) & r_{eh}(\varepsilon) \\ r_{he}(\varepsilon) & r_{hh}(\varepsilon) \end{pmatrix}$$

of excitations at energy $\varepsilon$ (of either an electron or a hole) approaching the junction from the normal metal. The tunneling barrier is described by a normal scattering matrix for electrons

$$S_I = \begin{pmatrix} \rho & \tau \\ \tau & \rho' \end{pmatrix}$$

with the transmission $|\tau|^2 = t$ and $\rho' = -\rho^* \tau / \tau^*$ assumed to be energy independent, and the corresponding matrix for holes is then just $S_I^*$. The N-S interface is described by the Andreev-reflection amplitude $a(\varepsilon)$ given by

$$a(\varepsilon) = \frac{1}{\Delta} \begin{cases} \varepsilon - sgn(\varepsilon)\sqrt{\varepsilon^2 - \Delta^2} & , |\varepsilon| > \Delta \\ \varepsilon - i\sqrt{\Delta^2 - \varepsilon^2} & , |\varepsilon| < \Delta \end{cases}.$$

The reflection amplitudes $r_{ee}(\varepsilon)$ and $r_{he}(\varepsilon)$ are found from an infinite series expansion considering all possible paths through which an incident electron is reflected as an electron or as a hole respectively. Taking the distance between the normal barrier and the S-N interface to zero, one obtains

$$r_{ee}(\varepsilon) = \rho + \tau a(\varepsilon) \rho'^* a(\varepsilon) \tau + \tau a(\varepsilon) \rho'^* a(\varepsilon) \rho' a(\varepsilon) \rho'^* a(\varepsilon) \tau + ... = \rho + \frac{\tau^2 \rho'^* a^2(\varepsilon)}{1 - |\rho|^2 a^2(\varepsilon)}$$

$$r_{he}(\varepsilon) = \tau a(\varepsilon) \tau^* + \tau a(\varepsilon) \rho'^* a(\varepsilon) \rho' a(\varepsilon) \tau^* + ... = \frac{|\tau|^2 a(\varepsilon)}{1 - |\rho|^2 a^2(\varepsilon)}$$

Defining $R_{ee}(\varepsilon) = |r_{ee}(\varepsilon)|^2$ and $R_{he}(\varepsilon) = |r_{he}(\varepsilon)|^2$, the current and current noise at zero temperature are then obtained by [2]

$$I = \frac{2e^2}{h} \int_0^{eV} d\varepsilon \left[1 - R_{ee}(\varepsilon) + R_{he}(\varepsilon)\right]$$

$$S = \frac{4e^2}{h} \int_0^{eV} d\varepsilon \left[R_{ee}(\varepsilon)\left[1 - R_{ee}(\varepsilon)\right] + R_{he}(\varepsilon)\left[1 - R_{he}(\varepsilon)\right] + 2R_{ee}(\varepsilon) R_{he}(\varepsilon)\right]$$

We calculated the ratio $S/2eI$ which does not show a dip around $eV = \Delta$ even though the current shows a peak from enhanced tunneling into the superconductor due to a singularity in the density of states.

## S2 - MBE Growth and sample fabrication

MBE Growth of InAs NWs.  The high-quality InAs NWs used in this study were grown by the Au-assisted, vapor-liquid-solid (VLS) method, in a high purity molecular beam epitaxy (MBE) system [7]. The epi-ready (111)B InAs substrate, glued onto a lapped silicon (Si) wafer, was initially heated in an 'introduction chamber' to 180°C for water desorption, followed by degassing at 350°C and a subsequent oxide blow-off with no intentional arsenic overpressure (in a dedicated treatment chamber attached to the MBE system). A thin layer of Au (less than 1$nm$ thick) was subsequently evaporated in the same chamber. Following transfer to the growth chamber, the substrate temperature was ramped up to ~550°C for ripening the Au layer into droplets with a rather uniform size and density distribution [7]. Lowering the growth temperature (to ~400°C), InAs growth was initiated with an $As_4$/In flux ratio of ~100, with resultant InAs NWs nucleating at the Au droplets and growing to a length of ~4-5μ$m$ with a diameter of 50-60$nm$. The NWs grow along the <0001> direction and have a pure Wurtzite structure mostly without any stacking faults (as verified by TEM imaging).

Device fabrication.  The sample was fabricated on a thermally oxidized Si/$SiO_2$ substrate (Si:$p^+$ doped and acts as a back gate). The NWs were detached from the growth surface by sonication, in ethanol and a droplet, later to dry, placed on a substrate with pre-arranged optical marks. Native oxide was removed and surface passivated with an ammonium polysulphide solution $(NH_4)_2S_x$=1:5, with the NWs immediately transferred into an evaporation chamber. Superconducting contacts, 5/120nm Ti/Al thick, were evaporated by electron beam evaporation.

# S3 – Detailed measurement setup

Figure S3 provides a detailed schematic diagram of the measurement setup. The experiment was performed in a dilution refrigerator, with an electron temperature of ~25mK, (inside the dashed region). The Josephson junction is voltage biased (with 5Ohms resistance at the Source), allowing access of all quiescent points in the *I-V* characteristic. Two electrical relays were employed, one at the input (at 300K) and one at the output (at 25mK); allowing switching from low frequency measurement (mode 1) - using the lock in technic, to a higher frequency (mode 2) - using a function generator and a spectrum analyzer. The actual measurements were done in mode 2, while measurements in mode 1 were performed in order to calibrate the higher frequency measurement.

*Mode* 1 *– low frequency measurement*

A calibration line allows calibrating the 5Ohms resistor after cooling. Applying DC voltage plus an AC signal and measuring the two-terminal AC current, allows calculating the static and dynamic conductance. The current was amplified by an external current amplifier [8], with $10^7$V/A conversion factor, followed by a DMM or a lock-in amplifier. The measured differential conductance was used to calibrate the higher frequency measurements.

*Mode* 2 *– higher frequency measurement*

While at DC the Drain is shorted through the coil *L*, the 600 kHz signal is divided between the junction resistance and $R_L$. The external voltage amplifier, SA-220F5, has a gain of 200, while the home-made 'cold' voltage preamplifier has a gain ~5. Noise measurements were performed by replacing the function generator (needed for the conductance measurements) with a DC source, and increasing the bandwidth of the spectrum analyzer. In our setup we also have two kinds of low pass filters, $LPF_1$ and $LPF_2$ which differ by their cut-off frequency. $LPF_1$ is placed both in RT and in base-temperature has a cutoff frequency of 80MHz (mini-circuit BHP-100+). $LPF_2$ is placed between them, also in base temperature and has a cut-off frequency of 2MHz.

## S4 - Estimating the background noise

The total voltage noise per unit frequency at the input of the 'cold' preamplifier:

$$S = S_{exc}r^2 + i_{amp}^2 r^2 + \upsilon_{amp}^2 + 4k_B Tr \quad \left[ V^2/Hz \right], \quad (S1)$$

where $S_{exc}$ is the excess current noise per unit frequency, $i_{amp}$ and $\upsilon_{amp}$ are the amplifier's current and voltage noises, respectively, $T$ is the temperature and $r$ is the resistance that the amplifier 'sees' at the resonance frequency (with a small frequency window):

$$r = R_{sample} \| R_L = \frac{R_{sample} R_L}{R_{sample} + R_L},$$

here $R_{sample}$ is the differential resistance of the sample and $R_L$ is the frequency independent load resistance. Note, that the $1/f$ contribution to the noise at $f_0$=600kHz is negligible. This is justified both from our measurements at high magnetic field as explained in S6 as well as from previous noise measurements done in our system to accurately extract integer and fractional charges of excitations in various 2DEG systems.

The background noise, subtracted from the total noise, is:

$$S_{BG} = i_{amp}^2 r^2 + \upsilon_{amp}^2 + 4k_B Tr \quad \left[ V^2/Hz \right]. \quad (S2)$$

Since the differential resistance is strongly dependent on biasing voltage $V_{SD}$, we first describe the procedure of determining the background noise. Since this noise (measured at zero bias) is laden with an emerging large Josephson current, it is quenched by applying a magnetic field stronger than $B_c$ ($B$~200mT), where the superconductivity is quenched. The differential conductance and the background noise were then measured as a function of the back-gate voltage, and thus as function of $r$, in the relevant range (Fig. S4). The values of the amplifier's noises obtained by fitting are in good agreements with the values we measure using other calibration methods. The electron temperature agrees well with that measured by other shot noise measurements.

## S5 – The critical magnetic field

In order to find the critical magnetic field, MAR conductance peaks are measured as a function of the magnetic field, with the spacing between the peaks directly proportional to the diminishing superconducting gap with magnetic field (Fig. S5).

## S6 – Number of conducting channels in the bare NW

Under high enough magnetic field the quantum charge passing the junction is that of the electron. The expression for shot noise provided in the text is that of a singly occupied spin-degenerate conducting channel,

$$S_{exc} = 2eI(1-t) \quad \text{and} \quad t = \frac{G}{g_Q} \quad , \quad \text{(S3)}$$

where $G$ is the conductance and $g_Q$ is the quantum of conductance $g_Q = 2\frac{e^2}{h}$.

The differential conductance and the $I\text{-}V_{SD}$ characteristic were measured after quenching superconductivity (but not lifting spin degeneracy) at the working voltage of the back-gate corresponding to the actual experiment (Figs. S6a and S6b). The noise is then measured as a function of $V_{SD}$ (blue curve in Fig. S6c), and the background noise is subtracted (red curve in Fig. S6c), and the excess noise is plotted in Fig. S6d. The theoretical curve, calculated using Eq. S4, plotted in a black dashed line, seems to agree nicely with the data. In order to test this further, we also plot the expected excess noise assuming two spin-degenerate channels, namely,

$$S_{exc} = 2eI_1(1-t_1) + 2eI_2(1-t_2) \quad \text{(S4)}$$

With $I_1$ and $I_2$ the current carried by each of the two channels, while $t_1$ and $t_2$ are the transmission of each of the two channels. If the total current, $I$, splits between the two channels in the following way

$$I_1 = \alpha I \qquad I_2 = (1-\alpha)I \quad , \quad \text{(S5)}$$

then,

$$t_1 = \alpha \frac{G}{g_Q} \qquad t_2 = (1-\alpha)\frac{G}{g_Q} \quad . \quad \text{(S6)}$$

Therefore, since we know *I* and *G*, we can plot $S_{exc}$ for a given α. In Fig. S6d we plot $S_{exc}$ for *α*=0.5 0.4 0.3 0.2 0.1 and 0.0. Note that *α*=0, being the single channel case, indeed fit best the data.

This measurement also allows us to show that the 1/f noise to the total noise is negligible. Since the 1/f noise is proportional to $I^2$, any non-negligible contribution of it would cause noise dependency on the current to deviate from formula S4 and to become non-linear. Since our measurement in Fig 6d is linear and completely coincides with the above formula we conclude that the 1/f noise is negligible.

## S7 – Nature of tunneling quasiparticles

Three possible models are suggested to account for the single quasiparticle tunneling taking place in the junction. We calculated the Fano-factor (*F*) for the models in order to see which one of them can account for the measured charge at the superconducting gap's edge – being smaller than *e* at a low transmission. In each table, we express the probability of an event to take place P(x) and its charge (X):

**Model 1:** Quasiparticles of charge *e* tunneling with probability *t*.

| x | P(x) |
|---|------|
| 0 | (1-t) |
| e | t |

$$\mu = et$$
$$\sigma^2 = e^2 t(1-t)$$
$$F = \frac{\sigma^2}{\mu} = e(1-t) \xrightarrow{t \ll 1} e$$

**Model 2:** Quasiparticles of charge *e\** tunneling with probability *t* and collapse as an electron with a probability *p* or as a hole with probability *q*, with *p+q*=1.

| x | P(x) |
|---|------|
| 0 | 1-t |
| e | tp |
| -e | tq |

$$\mu = et(p-q)$$
$$\sigma^2 = e^2 t - e^2 t^2 (p-q)^2$$
$$F = \frac{\sigma^2}{\mu} = e \frac{1-t(p-q)^2}{(p-q)} \xrightarrow{t \ll 1} e \frac{1}{(p-q)} > 1$$

**Model 3:** Quasiparticles of charge $e^*$ tunneling as a composite particle with probability $t$.

| x | P(x) |
|---|---|
| 0 | (1-t) |
| $e^*$ | t |

$$\mu = e^* t$$
$$\sigma^2 = e^{*2} t(1-t)$$
$$F = \frac{\sigma^2}{\mu} = e^*(1-t) \xrightarrow{t \ll 1} e^*$$

It is important to mention that the models above are considering only single quasiparticle tunneling across the junction neglecting higher order MAR contributions. When we lower the transmission we suppress the higher order MAR contributions and reveal a dip in the charge. This is clearly seen in our data as well as in the results of the theoretical model of **S1**. Once we suppressed these high order MAR contributions we observe a Fano factor which is smaller than $e$, which is only consistent with Model 3 above. In other words, the only way to observe a Fano factor that is lower than $e$ is both to suppress enough the high order MAR (going to low transmissions) as well as having a tunneling of quasiparticles carrying a fractional charge.

## S8 – Induced superconductivity on a single band

In this section we aim to support our claim in the manuscript for having a non-BCS density of state and specifically a sharper one in our 1D system.

Observing figure 3a in the main text it is possible to see negative differential conductance, this effect which is more apparent as the transmission is decreased originates as we will show from the change in the usual BCS density of states.
In figure S7 (a&b) we plot a measurement of the differential conductance as a function of the applied bias and the I-V curve. In figure S7 (c&d) we plot the theoretical predicted I-V curve and differential conductance based on the BCS density of state assuming a uniform transmission. The negative differential resistance which is clearly seen in the measurement and manifested in the experimental I-V curve as a peak in the current is not visible in the theoretical I-V, this suggests a different theoretical model should be given.

The origin of this discrepancy is the assumption of a linear dispersion which usually one considers in calculating the DOS. In a 1D wire, which has a parabolic dispersion, as the Fermi level is lowered to the bottom of the conduction band this assumption fails. Hence, the change in the DOS is more apparent as the fermi energy gets closer to the 'Van-Hove singularity'.

To show this we calculated the DOS and I-V curves as a function the fermi energy position. Figure 8(a, d and g) are the density of state for $E_F=5\Delta$, $2\Delta$ and $\Delta$ respectively. In figure 8(b, e and h) we plot each DOS when the fermi energy is defined as zero energy. It is already clear that as the Fermi level is pushed towards the bottom of the band the DOS is modified. In Fig 8(c, f and i) we calculate the I-V curves and show that the modified DOS gives rise to a peak in the I-V curve, similar to the one we showed in fig S7b.
In conclusion, the negative differential resistance which is seen in experiment as the device is pinched suggests a modification from the usual BCS density of state.

## **S9 – Charge partition in SIN junction**

In the SIS junction, the overlap between filled states of quasiparticles' wave-function and empty states (above the gap) in the two superconductors allows **tunneling** of quasiparticles with fractional charge. However, in the case of SIN, in the N side there are quasiparticles with charge *e* while in the S side there are quasiparticles with a **smaller charge**. Our physical picture suggests that tunneling of electrons, being of the higher charge is always dominant. In one polarity, the electrons that tunnel from N to S breaks to multiple quasiparticles; while in the opposite polarity, quasiparticles bunching to an electron (in N) takes place. This is similar to the known bunching in the $\nu = 1/3$ fractional quantum hall states where 3 quasiparticles, each with charge *e*/3, tunnel together to form an electron.

    Moreover, and in general, tunneling between two different materials, with different quasiparticles in each side, the current fluctuations will correspond to the larger charge transfer. For example, when the bias is smaller than Δ, electrons from the N region "bunch" to form Cooper pairs and the measured charge (via shot noise) is **2*e*** (via Andreev reflection).

# Figure Captions:

**Figure S1.** The Junction is modeled by a combination of a normal barrier and a perfect N-S interface. The calculation is done in the limit where the distance between the barrier and the N-S interface is zero.

**Figure S2.** SEM image of InAs NWs grown on (111)B InAs (micrograph taken at 45°). Note the uniformity of width and length of the NWs.

**Figure S3. Measurement setup.** A scanning electron micrograph of the device (scale bar, 200nm) connected to a detailed illustrated circuit.

**Figure S4. Background noise measurements.** (a) We start by measuring the differential conductance, $G_{sample}$, as a function of back gate voltage, $V_g$. The differential resistance is given by $R_{sample}=1/G_{sample}$ and is shown in (b). Then we calculate the resistance that the amplifier sees at its input, $R_{parallel}$, by taking $R_{sample}$ in parallel to $R_L$ and the result is shown in (c). (d) We then measure the background noise as a function of back gate voltage, $V_g$. This is done at magnetic field of 200mT to avoid effects related to superconductivity, and at zero bias to avoid any Shot noise. Combining the results allows us to plot the background noise as a function of $G_{sample}$ as shown in (e), or as a function of $R_{parallel}$ as shown in (f). In (f) we show the fit of the final result to a second order polynomial from which we obtain the coefficients of Eq. S2.

**Figure S5**. **Critical magnetic field of the Al contacts.** Differential conductance as a function of bias and magnetic field.

**Figure S6**. **Noise measurements at high magnetic field.** (a) Differential conductance vs. bias, $V_{SD}$, at magnetic field of 200mT. (b) *I-V* curve obtained by integrating the differential conductance. (c) Total voltage noise (in blue) and background noise (red) as a function of bias. (d) Excess current noise per unit frequency as s function of DC current through the device is plotted in blue. Theoretical lines of the expected excess noise assuming two channels carrying the current are plotted in dashed lines. The expected excess noise should follow $S_{exc} = 2eI_1(1-t_1) + 2eI_2(1-t_2)$, with $I_1 = \alpha I$ and $I_2 = (1-\alpha)I$ being the currents carried by each channel. Lines are plotted for *α*=0.5 0.4 0.3 0.2 0.1 and 0.0 (red to black) where the *α*=0 case reduces to the single channel scenario. The experimental data, falling on the *α*=0 line, leads us to conclude a single occupied channel.

**Figure S7**. **I-V curve of the experiment vs. BCS theory:** (a & b) Measurement of the differential resistance and I-V curves in very low transmission. (c & d) Differential resistance and I-V curves expected from BCS theory assuming a constant transmission in energy.

**Figure S8**. Each raw show the density of state and the I-V curve for a certain position of the fermi energy ($5\Delta$, $2\Delta$ and $\Delta$ above the minimum of the conduction band).

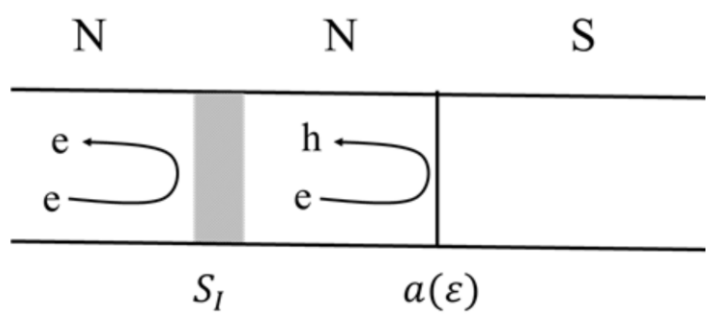

Fig. S1

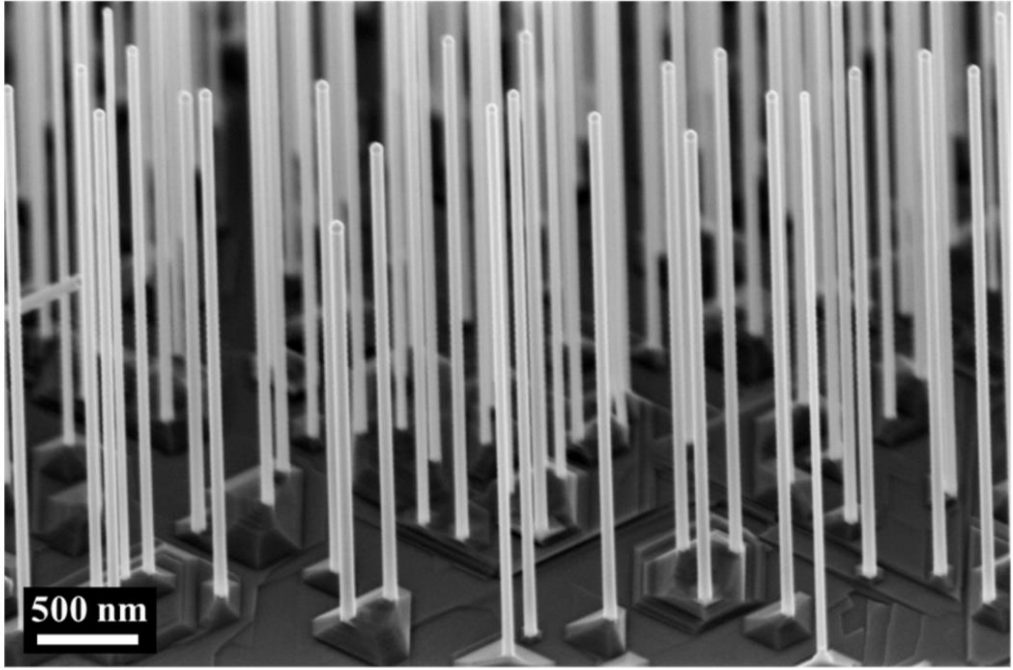

Fig. S2

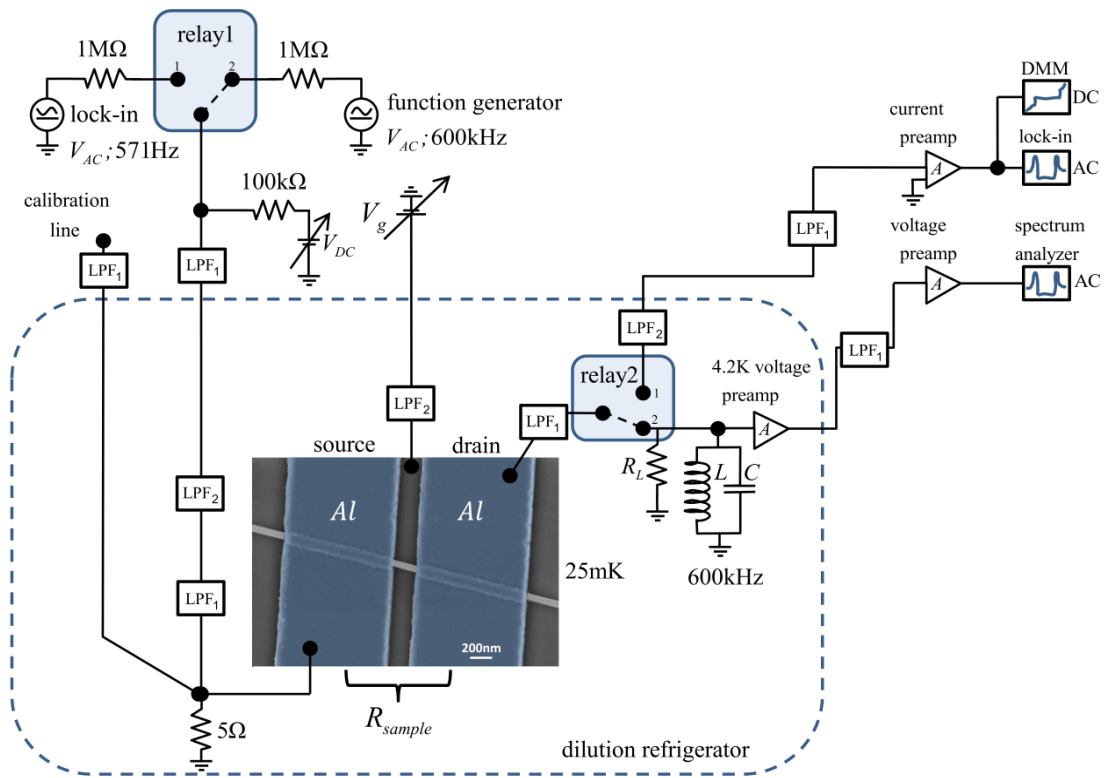

Fig. S3

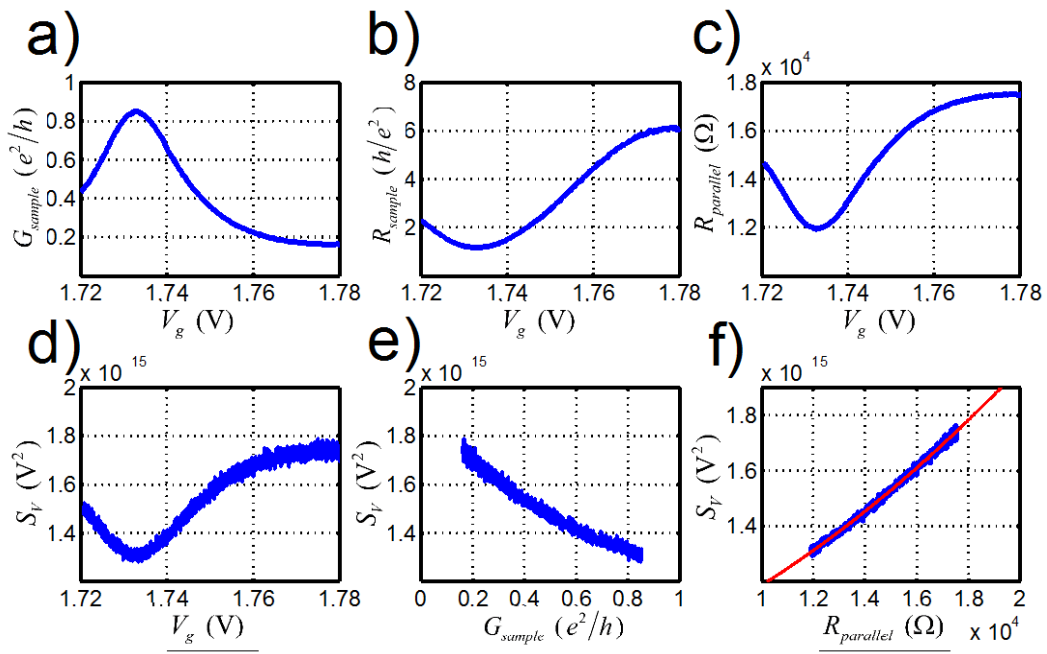

Fig. S4

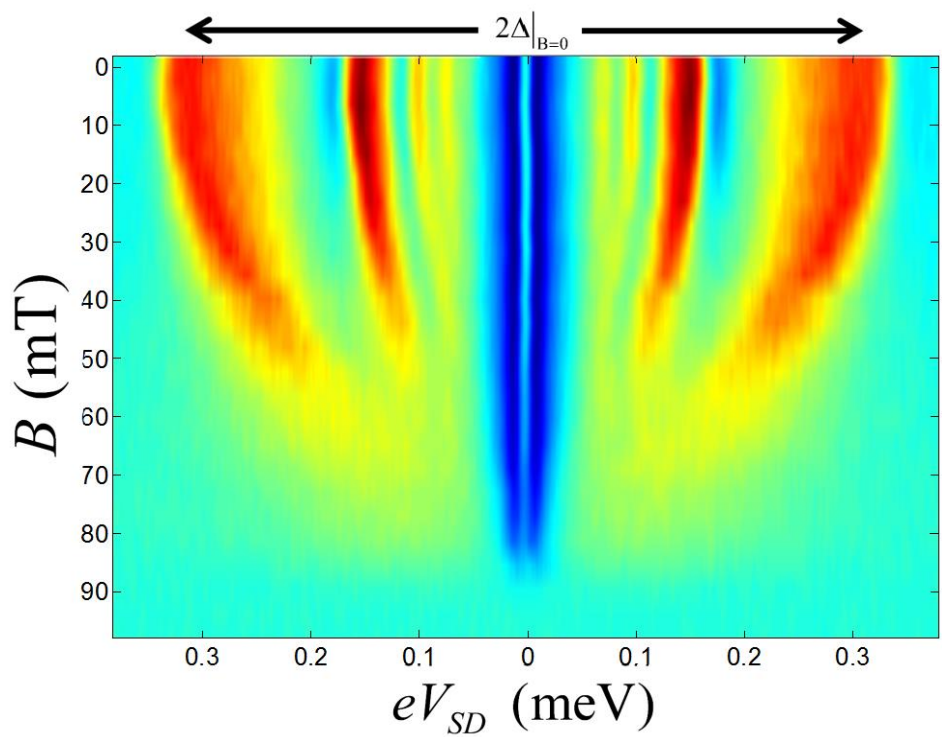

Fig. S5

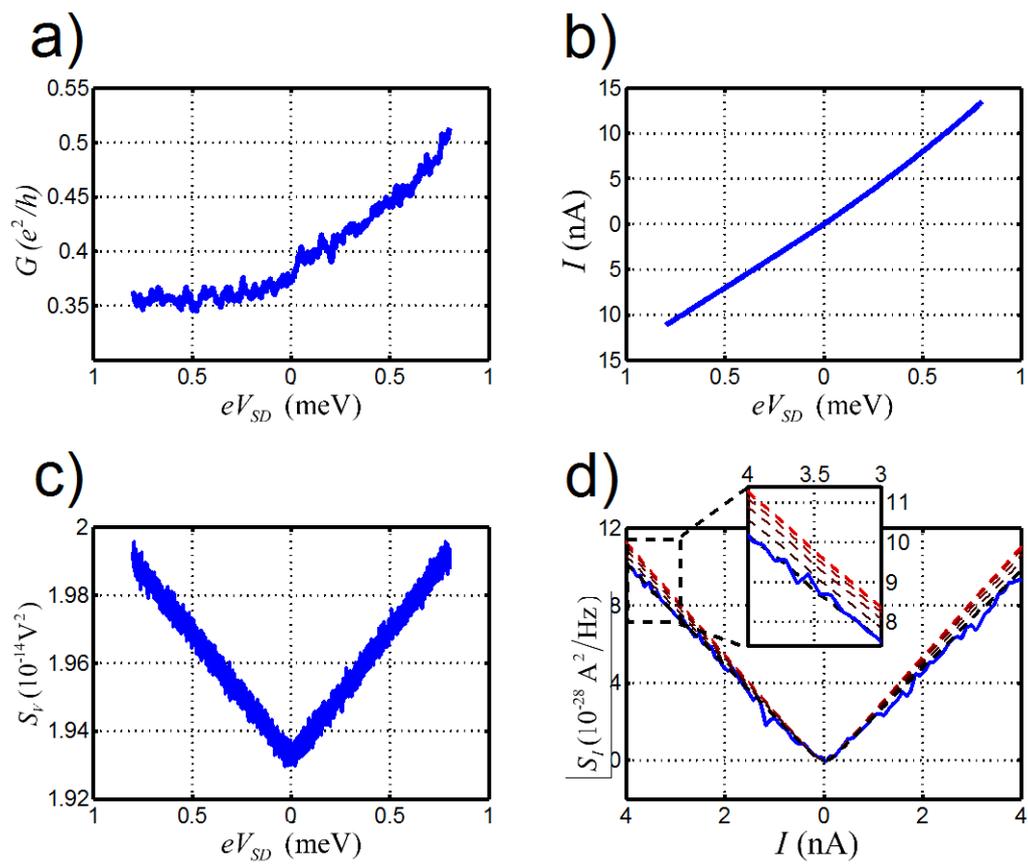

Fig. S6

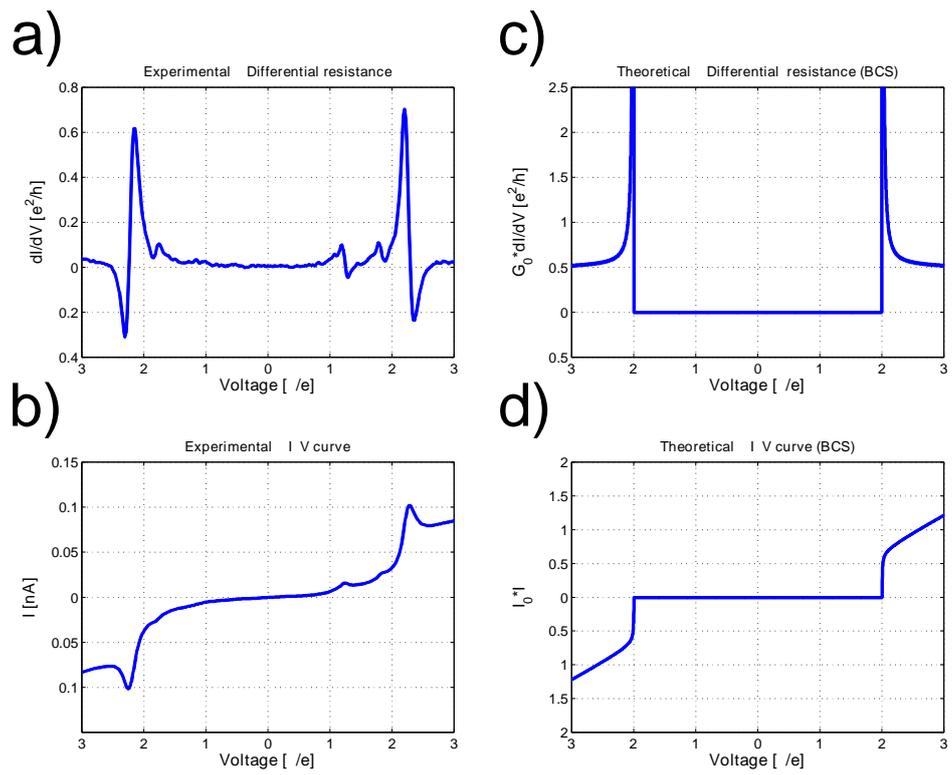

Fig. S7

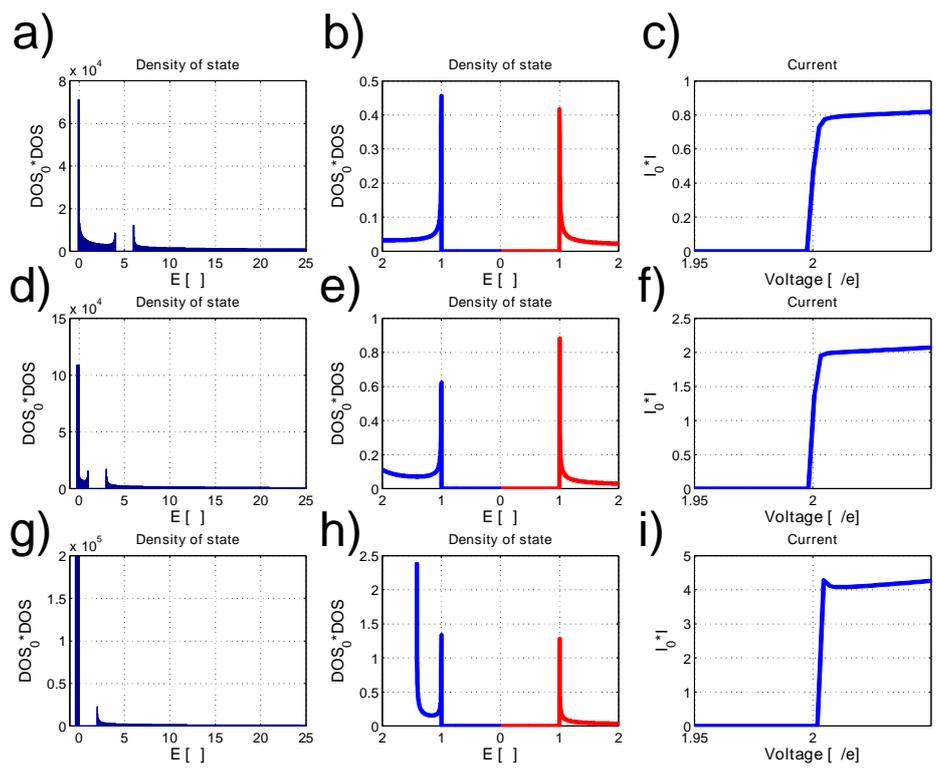

Fig. S8